\documentclass{PoS}

\newcommand{\gev}{\,{\rm GeV}}

\title{On the importance of gluon contributions to timelike and spacelike DVCS.}

\ShortTitle{On timelike and spacelike DVCS at NLO}

\author{H. Moutarde\\
Irfu-SPhN, CEA, Saclay, France}
\author{B. Pire\\
CPHT, {\'E}cole Polytechnique, CNRS, 91128 Palaiseau, France}
\author{F. Sabati\'e \\
Irfu-SPhN, CEA, Saclay, France}
\author{L. Szymanowski\\
National Center for Nuclear Research (NCBJ), Warsaw, Poland}
\author{\speaker{J. Wagner}\\
        National Center for Nuclear Research (NCBJ), Warsaw, Poland\\
        E-mail: \email{jwagner@fuw.edu.pl}}

\abstract{We emphasize how large $O(\alpha_s)$ corrections to timelike and spacelike virtual Compton scattering amplitudes are in the generalized Bjorken scaling regime, and in particular  the gluonic contributions, even in the medium energy range which will be studied intensely at JLab12 and in the COMPASS-II experiment at CERN. We stress that the timelike and spacelike cases are complementary and that their difference deserves much special attention.}

\FullConference{XXI International Workshop on Deep-Inelastic Scattering and Related Subject -DIS2013,\\
		22-26 April 2013\\
		Marseilles,France}

\begin{document}
\section{Introduction}
\begin{figure}[t]
\begin{center}
\includegraphics[width= 0.82\textwidth]{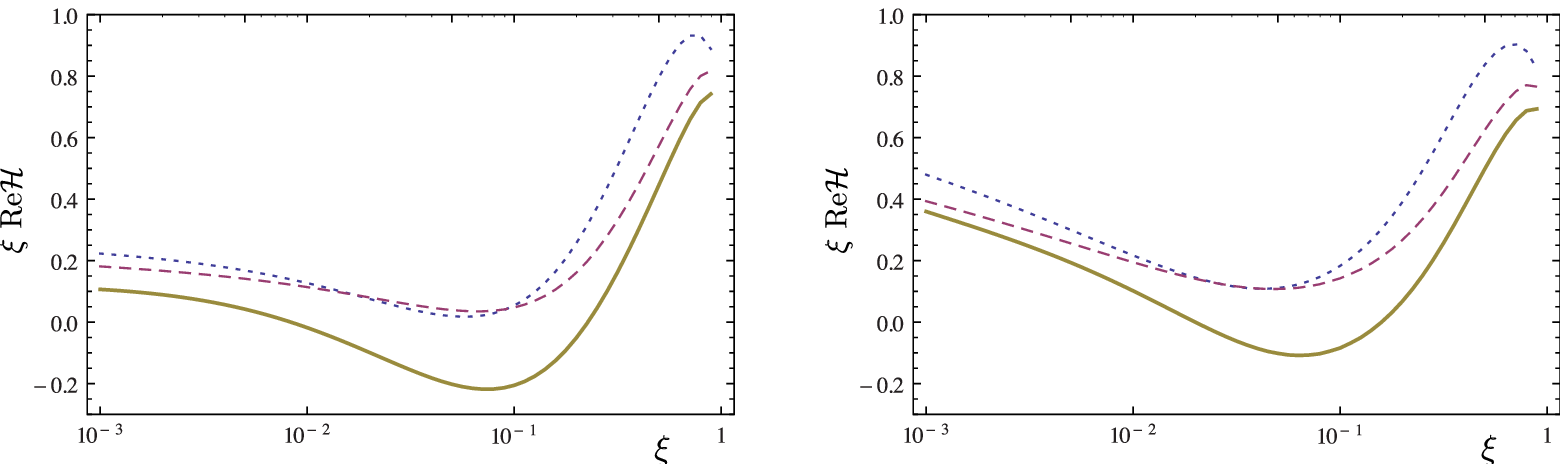} 
  \includegraphics[width= 0.82\textwidth]{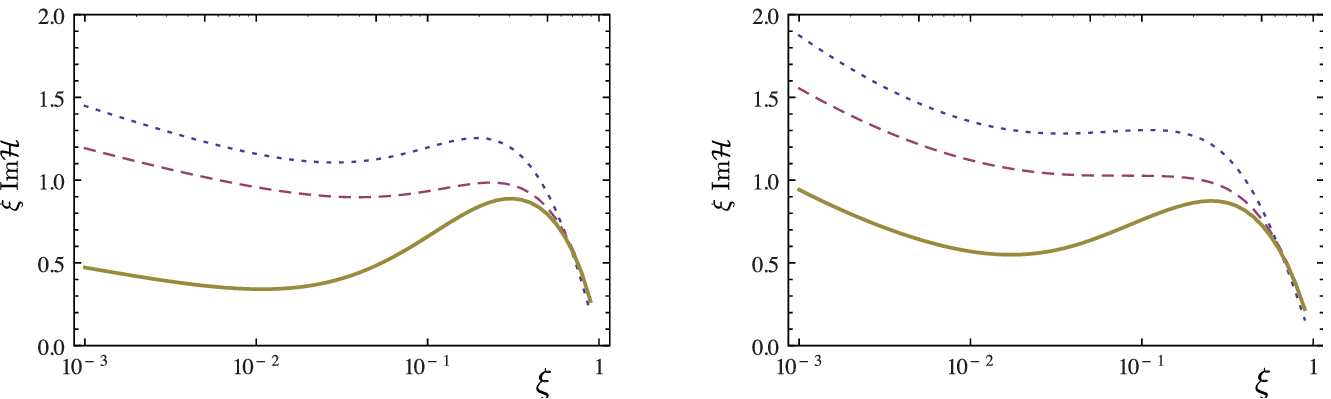} 
\caption{The real and imaginary parts of the {\it spacelike} Compton Form Factor $\mathcal{H}(\xi)$ multiplied by $\xi$, as a function of $\xi$ in the double distribution model based on Kroll-Goloskokov (left) and MSTW08 (right) parametrizations, for $
\mu_F^2=Q^2=4 \gev^2$ and $t= -0.1 \gev^2$, at the Born order (dotted line), including the NLO quark corrections (dashed line) and including both quark and gluon NLO corrections (solid line).}
\label{fig:DVCSRe2x2}
\end{center}
\end{figure}

We report on a recent work \cite{Moutarde:2013qs} where we studied $O(\alpha_s)$ corrections to spacelike Deeply Virtual Compton Scattering (DVCS) 
\begin{equation}
\gamma^*(q_{in}) N(P) \to \gamma(q_{out}) N'(P'=P+\Delta) \,,~~~~~q_{in}^2 =-Q^2,~~~~~q_{out}^2 =0\,,
\end{equation}
and to its crossed reaction, timelike Compton scattering (TCS)
\begin{equation}
\gamma(q_{in}) N(P)\to \gamma^*(q_{out}) N'(P'=P+\Delta)\,,~~~~~q_{in}^2 =0,~~~~~q_{out}^2 =Q^2\,,
\end{equation}
which are the model reactions to study generalized quark and gluon distributions (GPDs) which encode much information on the partonic content of nucleons. We denote the (positive) skewness variable as $\xi$ in the DVCS case, and as $\eta$ in the TCS case.
After factorization, the DVCS (and similarly TCS) amplitudes are written in terms of  Compton form factors  (CFF)$\mathcal{H}$, $\mathcal{E}$ and $\widetilde {\mathcal{H}}$, $\widetilde {\mathcal{E}}$ , as :
\begin{eqnarray}
\mathcal{A}^{\mu\nu}(\xi,t) &=& - e^2 \frac{1}{(P+P')^+}\, \bar{u}(P^{\prime}) 
\Big[\,
   g_T^{\mu\nu} \, \Big(
      {\mathcal{H}(\xi,t)} \, \gamma^+ +
      {\mathcal{E}(\xi,t)} \, \frac{i \sigma^{+\rho}\Delta_{\rho}}{2 M}
   \Big) \nonumber\\ &&\phantom{AAAAAAAAaaAA}
   +i\epsilon_T^{\mu\nu}\, \Big(
    {\widetilde{\mathcal{H}}(\xi,t)} \, \gamma^+\gamma_5 +
      {\widetilde{\mathcal{E}}(\xi,t)} \, \frac{\Delta^{+}\gamma_5}{2 M}
    \Big)
\,\Big] u(P) \, ,
\label{eq:amplCFF}
\end{eqnarray}
with the CFFs 
defined  as
\begin{eqnarray}
\mathcal{H}(\xi,t) &=& + \int_{-1}^1 dx \,
\left(\sum_q T^q(x,\xi)H^q(x,\xi,t)
 + T^g(x,\xi)H^g(x,\xi,t)\right) \; , \nonumber \\
\widetilde {\mathcal{H}}(\xi,t) &=& - \int_{-1}^1 dx \,
\left(\sum_q \widetilde {T}^q(x,\xi)\widetilde {H}^q(x,\xi,t) 
+\widetilde {T}^g(x,\xi)\widetilde {H}^g(x,\xi,t)\right).
\label{eq:CFF}
\end{eqnarray}

\section{How gluonic contributions affect Compton form factors}

Let us now present the results for spacelike and timelike Compton Form Factors (CFF), using the results of the NLO calculations of the coefficient functions which have been calculated in the DVCS case in the early days of GPD studies and more recently for  the TCS case \cite{PSW2}, the two results being simply related thanks to the analyticity (in $Q^2$) properties of the amplitude \cite{Muller:2012yq}:
\begin{eqnarray}
^{TCS}T(x,\eta) = \pm \left(^{DVCS}T(x,\xi=\eta) +  i \pi C_{coll}(x,\xi = \eta)\right)^* \,,
\label{eq:TCSvsDVCS}
\end{eqnarray}
where $+$~$(-)$ sign corresponds to vector (axial) case.

In our analysis we use two GPD models based on Double Distributions (DDs), as discussed in detail in Ref.  \cite{Moutarde:2013qs} : the Goloskokov-Kroll (GK) model \cite{GK3} and a model based on the MSTW08 PDF parametrization \cite{Martin:2009iq}. Our results are shown in Fig. 1- 2.  Comparing dashed and solid lines in the upper panels of both figures, one sees that gluonic contributions are so important that they even change the sign of the real part of the CFF, and are dominant for almost all values of the skewness parameter. A milder conclusion arises from a similar comparison of the lower panels; the gluonic contribution to  the imaginary part of the CFF remains sizeable for values of the skewness parameter up to $0.3$. 

\begin{figure}[t]
\begin{center}
  \includegraphics[width= 0.82\textwidth]{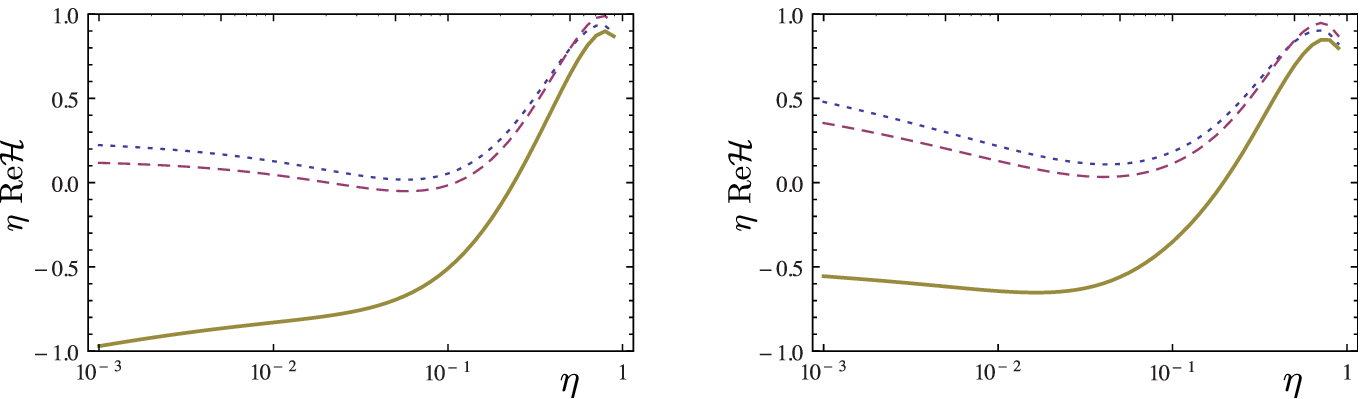} 
      \includegraphics[width= 0.82\textwidth]{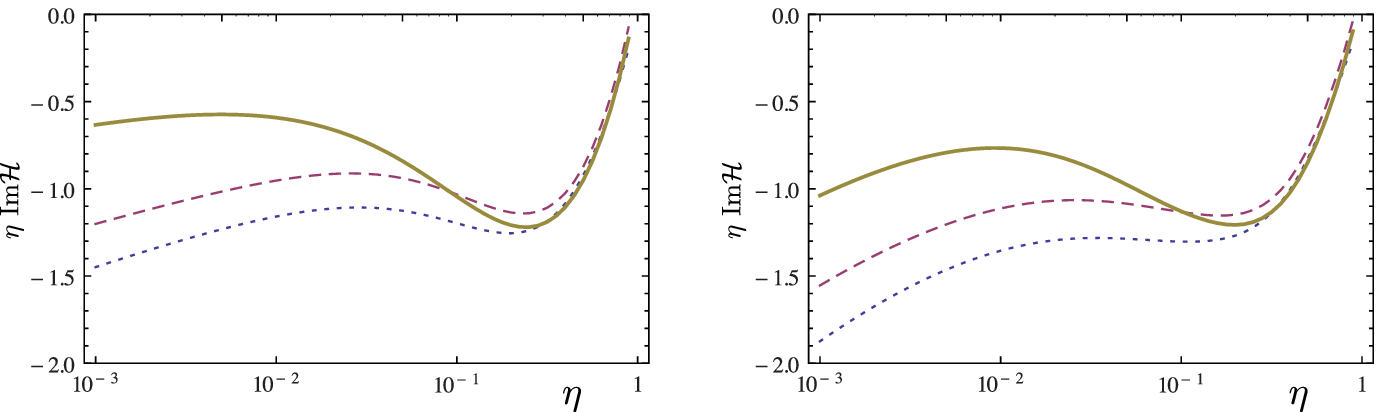} 
\caption{The real and imaginary parts of the {\it timelike} Compton Form Factor $\mathcal{H}$ multiplied by $\eta$, as a function of $\eta$ in the double distribution model based on GK (left) and MSTW08 (right) parametrizations, for $\mu_F^2=Q^2=4$~GeV$^2$ and $t=-0.1$~GeV$^2$.}
\label{fig:TCSRe2x2}
\end{center}
\end{figure}

\section{Gluonic effects on some DVCS and TCS observables}
\begin{figure}[t]
\begin{center}
  \includegraphics[width=8cm]{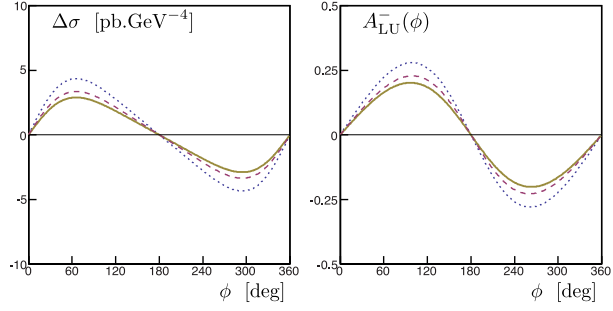} 
\caption{From left to right,  the difference of DVCS cross sections for opposite lepton helicities in pb/GeV$^4$, the corresponding asymmetry, all as a function of the usual $\phi $ angle   for $E_e=11 \gev, \mu_F^2=Q^2=4$~GeV$^2$ and $t= -0.2$~GeV$^2$. The GPD $H(x,\xi,t)$ is parametrized by the GK model. The contributions from other GPDs are not included. In all plots, the LO result is shown as the dotted line, the full NLO result by the solid line and the NLO result without the gluonic contribution as the dashed line. }
\label{fig:c1}
\end{center}
\end{figure}


\begin{figure}[hb!]
\begin{center}
  \includegraphics[width=12.5cm]{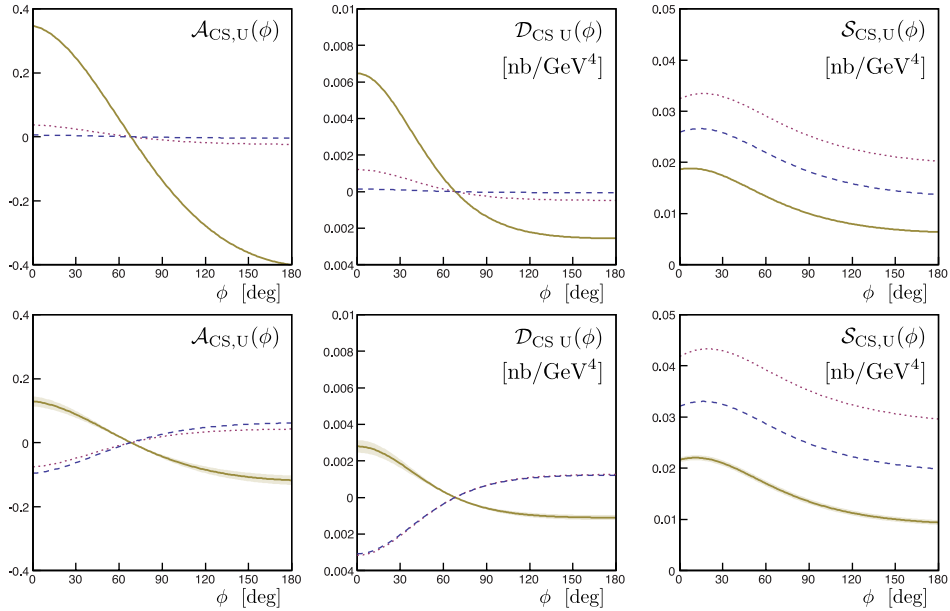}  
\caption{The DVCS observables for the COMPASS experiment, from left to right, mixed charge-spin asymmetry, mixed charge-spin difference and mixed charge-spin sum (in nb/GeV$^4$). The kinematical point is chosen as $ \xi=0.05, Q^2=4$~GeV$^2$, $t=-0.2$~GeV$^2$.  On the first line, the GPD $H(x,\xi,t)$ is parametrized by the GK model, on the second line  $H(x,\xi,t)$ is parametrized in the double distribution model based on the MSTW08 parametrization. The contributions from other GPDs are not included.}
\label{fig:compass}
\end{center}
\end{figure}

Let us now show the effects of the gluonic contributions to some of the DVCS and TCS observables at moderate energies. Fig. \ref{fig:c1} shows the difference and asymmetry for the lepton helicity dependent observables measured at JLab. The difference between the dotted and solid lines demonstrates that NLO contributions are important, whereas the difference between the dashed and solid lines shows that gluon contributions should not be forgotten even at low energy when a precise data set is analyzed. Fig. \ref{fig:compass}, which shows mixed charge spin observables  in COMPASS kinematics, magnifies the importance of gluonic contributions.

\begin{figure}[t!]
\begin{center}
  \includegraphics[width=6 cm]{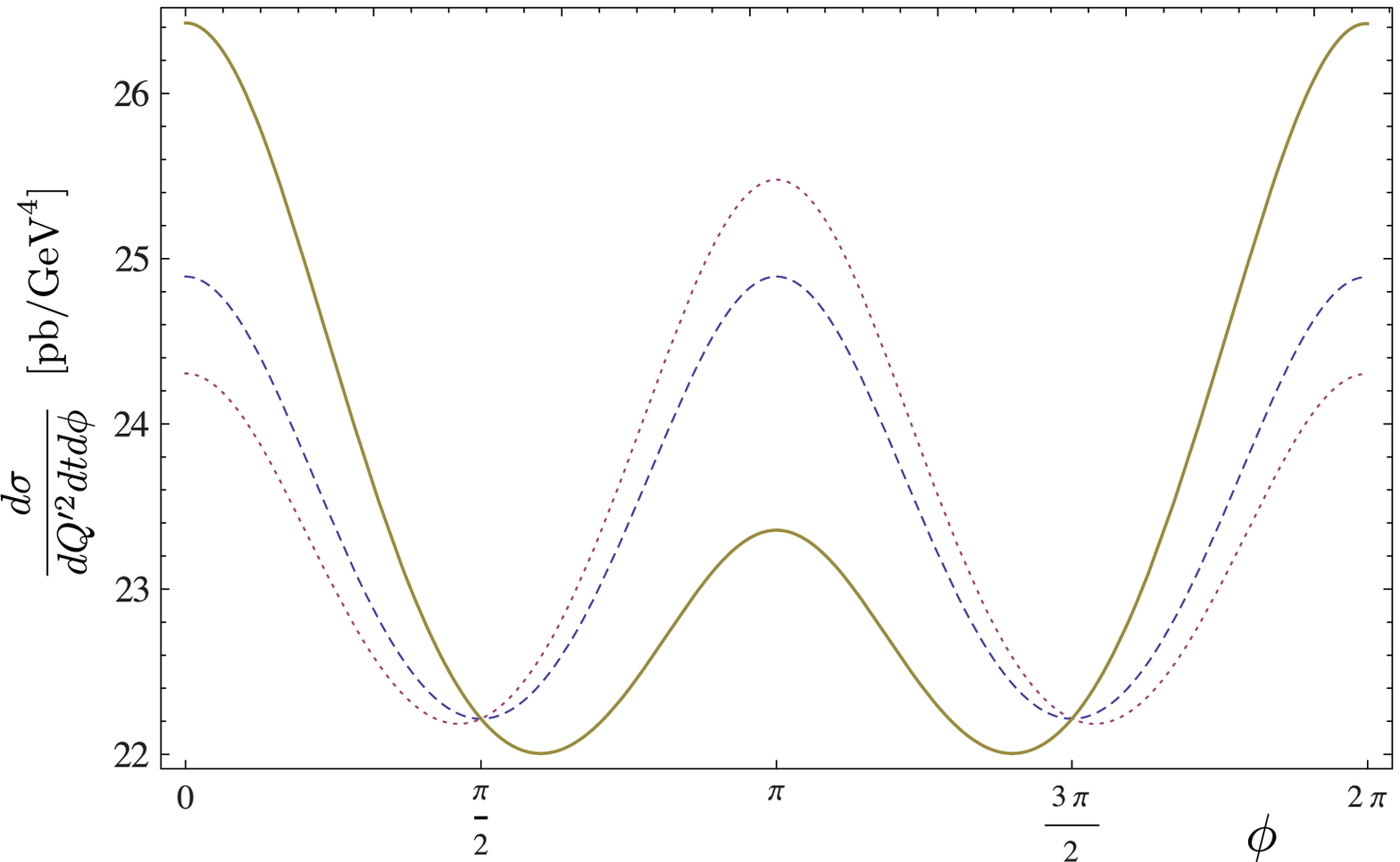}
  \hspace{1cm}
   \includegraphics[width=6 cm]{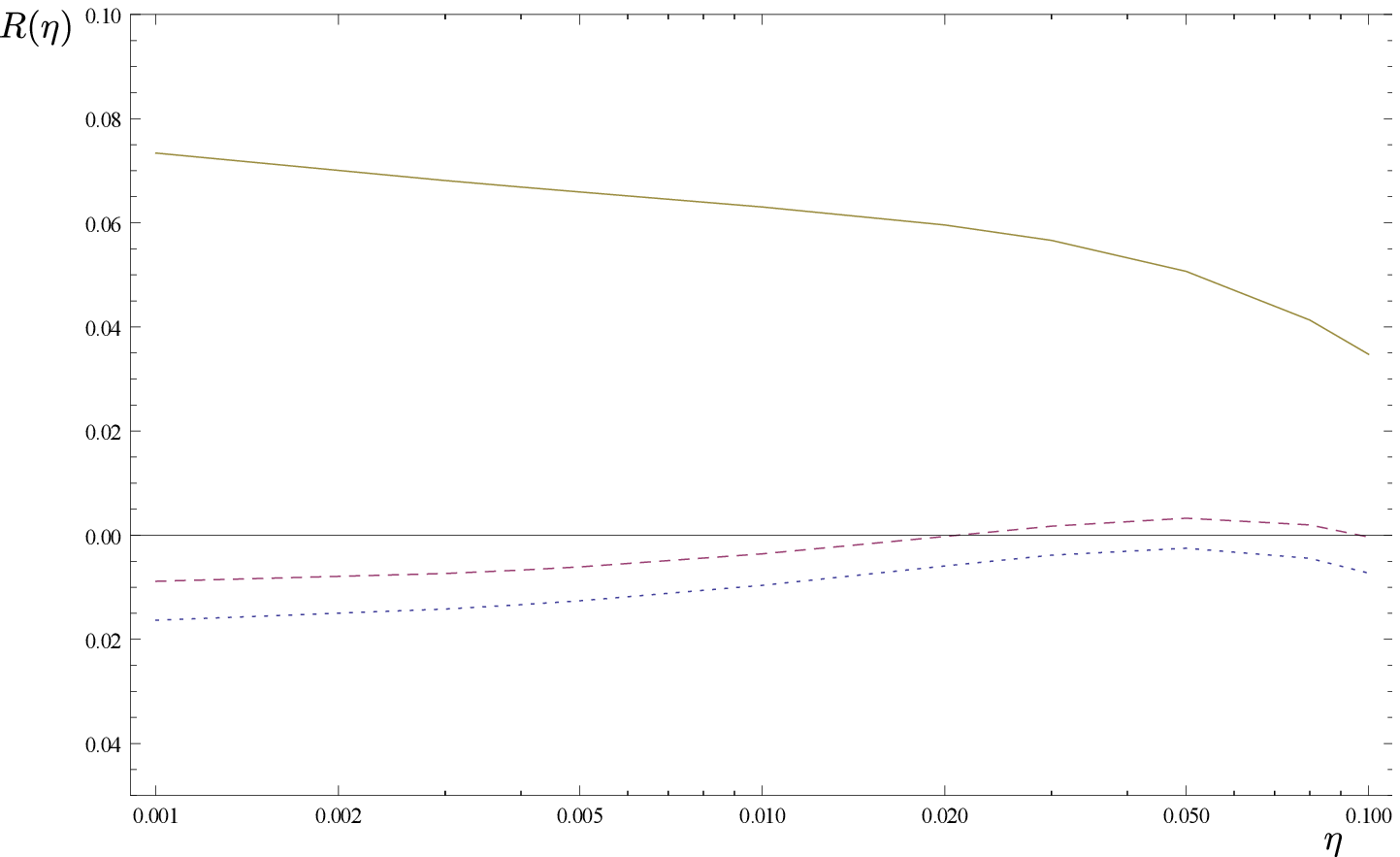}
\caption{(Left) The $\phi$ dependence of the cross-section at $E_\gamma = 10$ GeV,  $Q^2 =  \mu ^2 = 4$~GeV$^2$,  and $t= -0.1$~GeV$^2$ integrated over $\theta \in (\pi/4,3\pi/4)$: pure Bethe-Heitler contribution (dashed), Bethe-Heitler plus interference contribution at LO (dotted) and NLO (solid). (Right) Ratio R 
as a function of $\eta$, for $Q^2 = \mu_F^2 = 4$~GeV$^2$ and $t= -0.1$~GeV$^2$. The LO result is shown as the dotted line, the full NLO result by the solid line and the NLO result without the gluonic contribution as the dashed line.}
\label{fig:xsec_phidep}
\end{center}
\end{figure}


\begin{figure}[b!]
\begin{center}
  \includegraphics[width=0.4\textwidth]{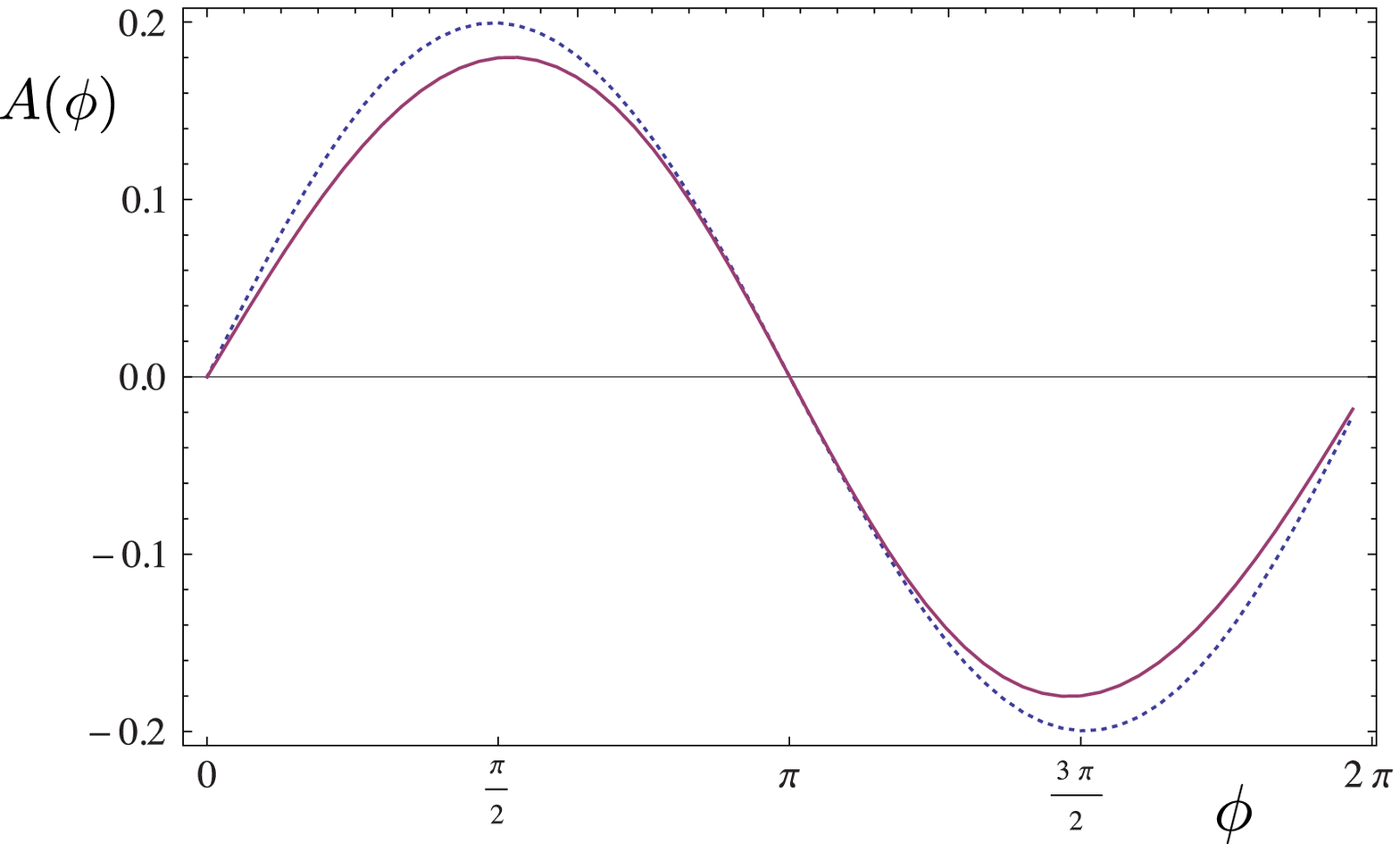}
  \hspace{0.05\textwidth}
  \includegraphics[width=0.4\textwidth]{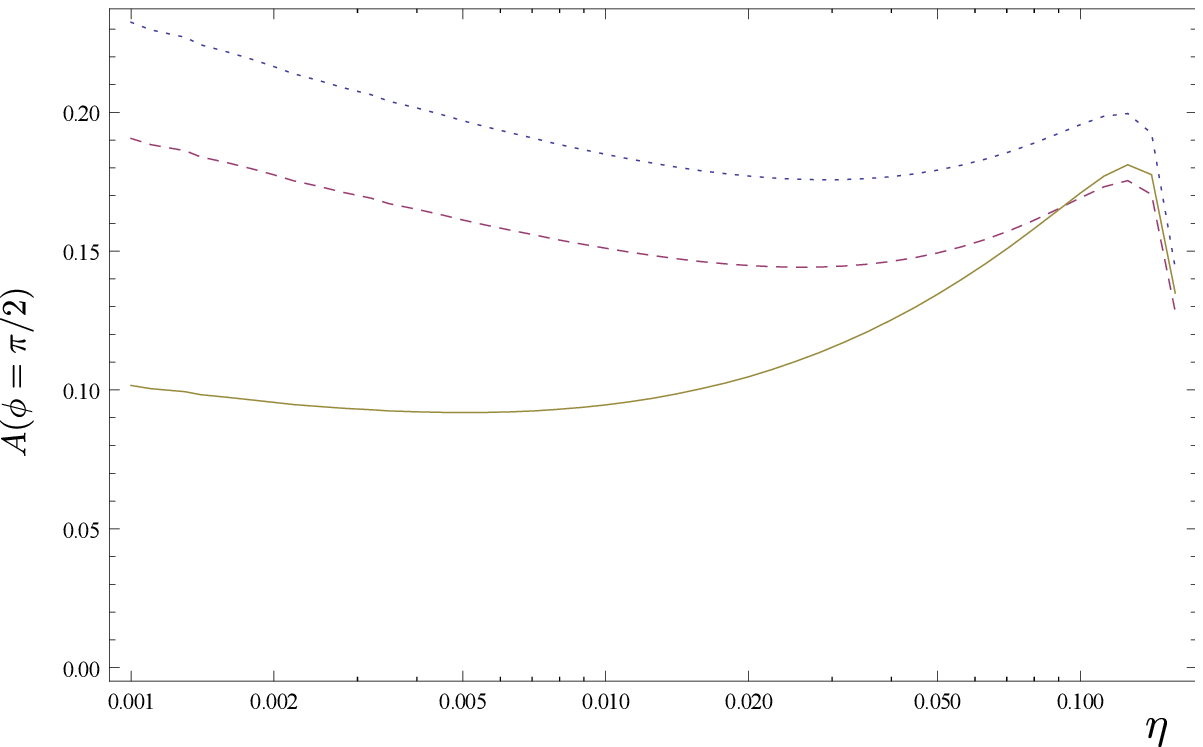}
\caption{
(Left) Photon beam  circular polarization asymmetry as a function of $\phi$, for $t=-0.1$~GeV$^2$, $Q^2 =  \mu^2 = 4$~GeV$^2$, integrated over $\theta \in (\pi/4,3\pi/4)$ and for $E_\gamma = 10$~GeV ($\eta \approx 0.11$).
(Right) The $\eta$ dependence of  the photon beam  circular polarization asymmetry for  $Q^2 =  \mu^2 = 4$~GeV$^2$,  and $t= -0.2$~GeV$^2$ integrated over $\theta \in (\pi/4,3\pi/4)$. The LO result is shown as the dotted line, the full NLO result by the solid line and the NLO result without the gluonic contribution as the dashed line.}
\label{fig:Asymmetry_xi}
\end{center}
\end{figure}


With respect to TCS, since the integrated cross section is dominated by the Bethe-Heitler process at low and medium energies, one needs to analyze differential observables to get information on the TCS amplitude. We show on Fig.~\ref{fig:xsec_phidep}a the azymuthal dependence for JLab kinematics; the difference between the dotted and solid lines demonstrates the importance of NLO effects on the TCS amplitude. On Fig.~\ref{fig:xsec_phidep}b we plot the ratio R, defined in Ref.~ \cite{Berger:2001xd} by
\begin{equation}
R(\eta) =  \frac{2\int_0^{2 \pi} d \varphi \,\cos \varphi\, \frac{dS}{dQ'^2dtd\varphi}}{\int_0^{2 \pi} d \varphi\frac{dS}{dQ'^2dtd\varphi}}
\,,
\label{eq:Rratio}
\end{equation}
where $S$ is a weighted cross section given by Eq.~(43) of Ref.~\cite{Berger:2001xd}. It is plotted as a function of the skewness $\eta$  for  $Q^2 =  \mu^2 = 4$~GeV$^2$,  and $t= -0.2$~GeV$^2$. In the leading twist the numerator is linear in the real part of the CFFs, and the denominator, for the kinematics we consider, is dominated by the Bethe - Heitler contribution. The inclusion of NLO corrections to the TCS amplitude is indeed dramatic for such an observable and includes also change of sign. As shown on Fig.~\ref{fig:xsec_phidep}b those corrections are dominated by the gluon contribution.

Imaginary parts of the CFFs are accessible through observables making use of photon circular polarizations \cite{Berger:2001xd}. The photon beam circular polarization asymmetry
\begin{equation}
A= \frac{\sigma^+ - \sigma^-}{\sigma^+ + \sigma^-}\,,
\end{equation}
is shown on the left part of Fig.~\ref{fig:Asymmetry_xi}, as a function of $\phi$ for the kinematical variables relevant for JLab: $Q^2 =4$~GeV$^2$= $\mu_F^2$, $t=-0.1$~GeV$^2$ and $E_\gamma = 10$~GeV (which corresponds to $\eta \approx 0.11$). The same quantity is shown on the right panel of Fig.~\ref{fig:Asymmetry_xi} as a function of $\eta$ for $\phi = \pi/2$ and $Q^2 =4$~GeV$^2$= $\mu_F^2$. The effect of the NLO corrections on that observable is rather large, ranging from $10\%$ at the $\eta=0.1$ (relevant for JLab) through $30\%$ at $\eta=0.05$ (relevant for COMPASS) up to $100\%$ at very small $\eta$'s. In that case the gluon contribution is important for $\eta<0.05$.

In conclusion, let us stress that our results point to the importance of understanding higher order effects, and maybe of resumming series of large contributions \cite{Altinoluk:2012nt}. They demonstrate that current attempts  \cite{Guidal:2009aa,Kroll:2012sm, Guidal:2013rya,  Kumericki:2009uq, Kumericki:2013br} to extract physics from DVCS data are but a first step in a long range program which will include data analysis from low to very high energy \cite{EIC}.

{\small
\section*{Acknowledgements} 
We are grateful to Markus Diehl, Dieter M\"{u}ller,  Stepan Stepanyan, Pawe{\l} Nadel-Turo\'nski and Samuel Wallon for useful discussions and correspondence. This work is partly supported by the Polish Grants NCN No DEC-2011/01/D/ST2/02069, by the Joint Research Activity "Study of Strongly Interacting Matter" (acronym HadronPhysics3, Grant Agreement n.283286) under the Seventh Framework Programme of the European Community, by the GDR 3034 PH-QCD, and the ANR-12-MONU-0008-01, and by the COPIN-IN2P3 Agreement.
}

\end{document}